\documentclass[pdflatex,sn-mathphys-num]{sn-jnl}

\usepackage{physics}
\usepackage{xcolor}
\usepackage{hyperref}
\usepackage{url}
\usepackage{graphicx}%
\usepackage{multirow}%
\usepackage{amsmath,amssymb,amsfonts}%
\usepackage{amsthm}%
\usepackage{mathrsfs}%
\usepackage[title]{appendix}%
\usepackage{xcolor}%
\usepackage{textcomp}%
\usepackage{manyfoot}%
\usepackage{booktabs}%
\usepackage{algorithm}%
\usepackage{algorithmicx}%
\usepackage{algpseudocode}%
\usepackage{listings}%

\theoremstyle{thmstyleone}%
\theoremstyle{thmstyletwo}%

\theoremstyle{thmstylethree}%

\begin{document}

\title[Dissipative adaptation in a driven spin-boson model within the path-integral formalism]{Dissipative adaptation in a driven spin-boson model within the path-integral formalism}


\author*[1,2]{\fnm{Elisa} \sur{I. Goettems}}\email{elisagtt@alumni.usp.br}

\author[1]{\fnm{Ricardo} \sur{J. S. Afonso}}
\author[1]{\fnm{Diogo} \sur{O. Soares-Pinto}}

\author*[3,4]{\fnm{Daniel} \sur{Valente}}\email{valente.daniel@gmail.com}

\affil[1]{\orgdiv{Instituto de Física de São Carlos}, \orgname{Universidade de São Paulo}, \orgaddress{\street{CP 369}, \city{São Carlos}, \postcode{13560-970}, \state{SP}, \country{Brazil}}}

\affil[2]{\orgdiv{Chemistry Department}, \orgname{Bar Ilan Univerity}, \orgaddress{\postcode{5290002}, \state{Ramat Gan}, \country{Israel}}}

\affil[3]{\orgdiv{Centro Brasileiro de Pesquisas F\'isica}, \orgaddress{\city{Rio de Janeiro}, \postcode{22290-180}, \state{RJ}, \country{Brazil}}}

\affil[4]{\orgdiv{Instituto de F\'isica}, \orgname{Universidade Federal de Mato Grosso}, \orgaddress{\city{Cuiab\'a}, \postcode{78060-900}, \state{MT}, \country{Brazil}}}


\abstract{We investigate the dissipative adaptation hypothesis in a quantum regime using a system-reservoir approach. This hypothesis proposes that self-organization arises from a system's ability to dissipate the work transiently absorbed from an external drive. 
We analyze the quantum dynamics of a driven open system described by a time-dependent spin-boson Hamiltonian modeling a particle in a metastable double-well potential with controllable asymmetry. We explore how the work provided by the dynamic potential is related to the transition probability between the two ground states of the double well. These studies motivate further investigations of the driven spin-boson model toward an understanding of the system's evolution and its thermodynamic implications. }


\maketitle

\section{Introduction}\label{sec:introduction}

In equilibrium statistical mechanics, the probability of observing a given microscopic configuration is determined by its energy through the Boltzmann distribution. 
In contrast, when a system is driven away from equilibrium by external forces that perform work over finite times, no analogous universal distribution exists. 
In such nonequilibrium settings, relating the likelihood of a given outcome to thermodynamic quantities, such as work or heat, becomes a fundamental challenge.

Developing a general thermodynamic understanding of driven nonequilibrium systems is particularly compelling in light of biological phenomena. 
Living organisms operate far from equilibrium and maintain organized structures through continuous energy exchange with their environment. 
Since equilibrium thermodynamics alone cannot account for such behavior, nonequilibrium frameworks are important for elucidating the physical principles underlying life-like self-organization and driven assembly processes.
Motivated by this perspective, England proposes the dissipative adaptation hypothesis \cite{england2015dissipative}, which generalizes the Boltzmann distribution to driven nonequilibrium systems. 
This hypothesis states that during nonequilibrium evolution, the probability of a system transitioning between macrostates depends not only on the energies of the initial and final states but also on the history of work absorption and heat dissipation along the trajectory. 
In this view, self-organization emerges as a thermodynamic consequence of enhanced dissipation under external drivings.
The idea draws parallels with the understanding of evolutionary adaptation in biological contexts \cite{perunov2016statistical}, i.e., structures formed far from equilibrium may appear adapted in the sense that they are especially effective at either absorbing or avoiding the absorption of work from the driving environment \cite{prlengland,commphysdv}.
Such behavior has been demonstrated in classical driven systems and has stimulated renewed interest in the thermodynamics of self-organization \cite{prekondepudi,natnanohuck}.

Although the hypothesis of dissipative adaptation has been mainly explored in classical contexts, its extension to quantum systems is still under investigation \cite{qda,valente21,qdaN,pre2025}. 
Understanding how quantum coherence, dissipation, and external driving influence adaptive behavior remains an open problem in the thermodynamics of nonequilibrium quantum systems.
From a practical perspective, this suggests the search for other regimes for quantum (or perhaps semiclassical) information processing by employing open quantum systems at finite temperatures as an alternative to ever better isolation.

Here, we investigate the dissipative adaptation hypothesis in a quantum regime using the driven spin-boson model.
This model represents a quantum particle that can coherently jump between the two minima of a bistable potential, and undergoes dissipative effects from a thermal bath of harmonic oscillators.
Beyond the presence of bistability and dissipation in this model, our choice is motivated by the fact that it describes the physics of superconducting qubits extensively used in quantum information processing setups \cite{leggett1984quantum,leggett1987dynamics,devoret,devoret2,caldeiraRBEF}.
By employing the path-integral formalism, we analyze the dynamics of a quantum system coupled to a thermal reservoir and subjected to time-dependent driving.
We find a direct relationship between the transition probability of the particle from one minimum to another in the double well at a finite time and the work transiently absorbed from the drive during that transition.

The article is organized as follows. 
In Sec II, we review the hypothesis of dissipative adaptation, summarizing its classical formulation and discussing existing extension to the quantum regime. 
In Sec. III, we introduce the driven spin-boson model and describe the system-reservoir framework used to characterize the nonequilibrium dynamics.
In Sec. IV, we formulate quantum work within the path integral framework. 
In Sec. V, we present our results on the relation between transition probabilities and work.
Finally, Sec. VI summarizes our results and outlines directions for future investigations.

\section{Dissipative Adaptation Hypothesis}\label{sec:hypothesis_dissipative_adaptation}
\subsection{The classical framework}
The thermodynamics of nonequilibrium systems has been extensively explored in the past few years~\cite{le2004equilibrium,lebon2008understanding,de2013non}. 
In this context, the dissipative adaptation hypothesis, introduced by England~\cite{england2013statistical,england2015dissipative}, represents an important step towards extending concepts from equilibrium statistics, such as the Gibbs distribution, to systems far from equilibrium.
As stated above, England formalized this idea by focusing not on the occupation probabilities of single states, but rather on the probabilities of trajectories connecting initial and final configurations. 
Using fluctuation relations such as the Jarzynski equality \cite{jarzynski} and the Crook's theorem \cite{crooks1999entropy}, the following relation between the probabilities $p_{i\rightarrow j,k}(t)$ of transitions from an initial state $i$ to distinct final states $j$ and $k$, over a finite time interval $t$, has been derived~\cite{england2015dissipative,perunov2016statistical}
\begin{align}\label{eq:england}
    \frac{p_{i\rightarrow j}(t)}{p_{i\rightarrow k}(t)}
    =
    \exp{-\frac{\Delta E_{jk}}{k_B T}}
    \frac{p_{j^*\rightarrow i^*}^{*}(t)}{p^{*}_{k^{*}\rightarrow i^{*}}(t)}
    \frac{\langle e^{-W/k_B T}\rangle_{i\rightarrow k}}{\langle e^{-W/k_B T} \rangle_{i\rightarrow j}},
\end{align}
where $T$ is the environment temperature, $k_B$ is Boltzmann's constant, and $W$ is the stochastic work absorbed in the trajectory. 
The average is taken over the trajectory ensembles between initial and final states.
This expression highlights three distinct contributions to nonequilibrium evolution~\cite{perunov2016statistical}
\begin{enumerate}
    \item an equilibrium-like Boltzmann factor, which depends on the energy difference $\Delta E_{jk} = E_j-E_k$ between two final states. 
    \item kinetic accessibility over time, as quantified by time-reversal probabilities $p^*$, indicates that the heights of the barriers impose restrictions both in the forward and in the reversed senses of thermal-equilibrium stochastic dynamics, and
    \item most importantly, a work-dependent quantity encoding how transient energy absorption along trajectories may favor specific thermally inaccessible states, thus leading to broken detailed balance.
\end{enumerate}
In driven systems, the third term can dominate, favoring transitions that absorb work reliably with small fluctuations. 
In this sense, self-organization emerges as a thermodynamic consequence of dissipation: 
more persistent states are those with a history of enhanced work absorption followed by heat dissipation. 
This provides a nonequilibrium generalization of the Boltzmann distribution, 
applicable to systems evolving over time under external driving.
Irreversibility plays an important role in this framework. 
When absorbed energy is subsequently dissipated into the environment, backward transitions become suppressed, and the system loses memory of its initial configuration. 
As a result, symmetry breaking and localization can arise.
\subsection{A quantum version}\label{sec:quantum_dissipative}
When studying dissipative adaptation in a quantum regime, two main challenges arise.
First, in the low-temperature limit $T\rightarrow 0$, quantum fluctuations become dominant.
At $T=0$, the classical formulation no longer applies.
Second, the definition of work in open quantum systems is not uniquely defined. 

Despite these difficulties, recent studies have begun to explore quantum manifestations of dissipative adaptation in quantum regimes. 
Refs. \cite{qda,valente2021quantum,qdaN} have studied three-level systems in $\Lambda$ configuration coupled via electric dipole interaction to a quantized electromagnetic environment at zero temperature ($T=0$). 
In their model, a single photon pulse acts as a nonequilibrium work source, possibly inducing transitions between the two ground states, $\ket{1}$ and $\ket{2}$, through an excited state $\ket{3}$ serving as an energy barrier.

By solving exactly the full system-environment dynamics, the authors have derived a relation between transition probabilities and the average absorbed work, 
\begin{align}\label{eq:daniel}
    p_{1\rightarrow 2} (\infty) &= \frac{\Gamma_2}{\Gamma_1+\Gamma_2}\frac{\langle{W_\mathrm{abs} \rangle}}{\hbar\omega_L},\\
    p^{*}_{2\rightarrow 1} (t)&=0.
\end{align}
where the vanishing backward probability reflects the irreversibility of the driven process at zero temperature, and the average work 
$\langle W_\mathrm{abs} \rangle$
performed by the photon (of frequency $\omega_L$) on the system generates nonequilibrium dynamics.
The average work in this case can be consistently defined both in Schr\"odinger and in Heisenberg pictures, as shown in Refs.\cite{qda,valente21}.
Kinetic accessibility appears in the spontaneous emission rates $\Gamma_{1,2}$ from the excited state.
Curiously, in this model the transition probability depends linearly, not exponentially, on the absorbed work.

These results motivate the exploration of more diverse settings.
Here, we address the spin-boson model, as it is particularly useful in describing superconducting qubits. 
From a more fundamental perspective, the two-level structure of the spin-boson model makes us question whether the quantum dissipative adaptation shown above survives without the third (higher energy) level of the $\Lambda$ model, and in the presence of quantum tunneling between the two ground states of a bistable potential.
Also, the path-integral approach to the driven spin-boson model allows us to derive finite temperature results.

\section{Driven spin-boson model}
\label{chap:spin_boson}
The spin-boson model is a type of Caldeira-Leggett model \cite{caldeira1981influence,caldeira1983quantum,caldeira1983path} that describes a two-level system coupled to a bosonic environment. Physically, it models a particle tunneling between two quasi-degenerate minima of a double-well potential while interacting with a bath of harmonic oscillators that are sensitive to the position of the particle inside the double well. 
When the quantum particle becomes delocalized by means of quantum tunneling, the bath induces decoherence and energy dissipation~\cite{leppakangas2018quantum,leggett1984quantum,caldeira2014introduction}. 

\begin{figure}[htb]
    \centering
    \includegraphics[scale=1.2]{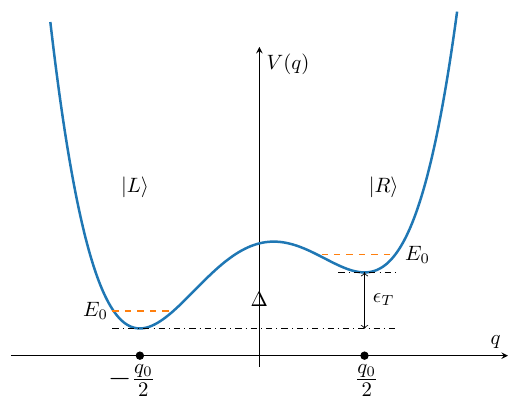}
    \caption{Asymmetric double well potential describing a dissipative two-level system. The minima are located at $q=\pm q_0/2$, corresponding to the localized states $\ket{L}$ (left) and $\ket{R}$ (right), respectively. The asymmetry parameter $\epsilon$ defines the energy bias between the two wells. The local harmonic frequencies around each minimum are denoted by $\omega_{-}$ (left well) and $\omega_{+}$ (right well). 
    $\epsilon_T<0$: the left well is energetically favored (stable configuration), and a particle initially localized in $\ket{L}$ starts in the global minimum. 
    If $\epsilon_T>0$: the left well becomes metastable, and a particle initially in $\ket{L}$ starts in a higher-energy local minimum.}
    \label{fig:poço}
\end{figure}

We employ the driven spin-boson model to investigate the relation between transition probabilities and work absorption in a nonequilibrium quantum setting. 
The model provides a minimal yet realistic platform for studying dissipative adaptation, as it incorporates tunneling, dissipation, and externally controlled driving (Fig. \ref{fig:poço})
\cite{orth2013nonperturbative,funo2018path}. 

The total Hamiltonian is written as 
\begin{align}\label{eq:biased}
    H_S&=-\frac{\hbar \Delta}{2}\sigma_x-\frac{\hbar\epsilon_T(t)}{2}\sigma_z\\ \nonumber
    H_I &= \frac{q_0\sigma_z}{2}\sum_k C_k x_k\\ \nonumber
    H_{R} &= \sum_k \left(\frac{m_k \omega_k x_k^2}{2}+\frac{p_k^2}{2m_k}\right).
\end{align}
where $\sigma_x = \ket{L}\bra{R}+\ket{R}\bra{L}$, and $\sigma_z = \ket{R}\bra{R} - \ket{L}\bra{L}$, are Pauli matrices acting on the two-level system.
State $\ket{L}$ ($\ket{R}$) corresponds to a particle localized on the left (right) side of a double well.
The tunneling rate is given by $\Delta$, and $\hbar \epsilon_T(t)=\epsilon_0 + \epsilon_d(t)$ gives the instantaneous energy difference between the two minima.
In the absence of driving ($\epsilon_d(t) = 0$), the Hamiltonian describes a static double-well potential. 
At equilibrium, the populations of the two localized states are expected to obey the Gibbs distribution among the eigenstates of $H_S$.
A time-dependent bias might break this, allowing for a controlled population transfer between the wells.
The reservoir is modeled as a collection of harmonic oscillators with coordinates $x_k$, momenta $\;p_k$, masses $\;m_k,$, and frequencies $\omega_k$.
These can be thought of as phonon frequency modes of the material surrounding the particle of interest.
Parameter $q_0$ denotes the spatial separation between the potential minima, and $C_k$ characterizes the system-bath coupling strength. 
Depending on the side the particle occupies, each oscillator is pushed or pulled, with strength $\pm q_0 C_k /2$.
In the continuum limit ($\sum_k \rightarrow \int dk$), the reservoir is characterized by the spectral function
$J(\omega)=\pi \sum_k (C_k^2/2m_k \omega_k)\delta(\omega-\omega_k)$.
Memory effects (non-Markovianity) can be taken into account by the appropriate modeling of $J(\omega)$.
Typically, one assumes that
$J(\omega)=2\pi \alpha \omega^s e^{-\omega/\omega_c}$,
where $\alpha\geq 0$ is the dimensionless dissipation strength 
($\alpha \equiv \eta q_0^2/2\pi \hbar$), 
and $\omega_c$ is a high-frequency cutoff ($\omega_c^{-1}$ gives the reservoir correlations timescale).
In the case of an Ohmic environment, one sets $s=1$.

We consider an initial state localized in the left well as shown in Fig.\ref{fig:poço}, 
\begin{align}
\rho_S(0) = \ket{L}\bra{L},
\end{align}
while the reservoir is prepared in a thermal Gibbs state, $\rho_{R}(0) = e^{-H_{R}/k_B T}/\mbox{Tr}(e^{-H_{R}/k_B T})$.
The total initial density matrix is thus factorized,
$\rho(0)=\rho_S(0) \rho_R(0)$,
corresponding to an initially equilibrated system-environment configuration prior to the application of the external drive.
The reduced density matrix of the system, $\rho_S(t)=\mbox{Tr}_R\rho(t)$, is obtained using the Feynman-Vernon influence functional formalism. 
See details in Appendix \ref{appendix:path}. 

The transition probability $p_{L\rightarrow R}(t)$ from $\ket{L}$ to $\ket{R}$ at time $t$ is analytically obtained
after summing over all possible sequences of tunneling events (blips) and localization intervals (sojourns), 
reading \cite{grifoni1995cooperative,grifoni1993nonlinear,grifoni1998driven}
\begin{align}\label{eq:trans}
    p_{L\rightarrow R}(t) =-\sum_{n=1}^{\infty}\left(-\frac{\Delta^2}{4}\right)^{n}\int_{0}^{t}\mathcal{D}_n\{t_j\}
    \sum_{\{\xi_j\}}
    Q_n e^{i\Phi_n}
    \sum_{\{ \eta_k \}'} H_n.
\end{align} 
Here, time integration is denoted as
\begin{align}
\int_{0}^{t}\mathcal{D}_n\{t_j\} = \int_0^t dt_{2n} \int_0^{t_{2n}} dt_{2n-1} ... \int_0^{t_2}dt_1.
\end{align}
The summation indices span 
$\xi_j = \pm 1$ and $\eta_j = \pm 1$.
The prime in $\{ \eta_k \}'$ denotes restriction
$\eta_0 = -1$ (for the initial state $L$).
These numbers correspond to the discretization of the stochastic position that jumps between the two minima so that $q(t') = \pm q_0/2$.
One thus defines
$\xi(t') \equiv [q(t') - q'(t')]/q_0$ 
to describe the off-diagonal part (the quantum coherent delocalization of the wavepacket)
and
$\eta(t') \equiv [q(t')+q'(t')]/q_0$,
to the diagonal part (the ``center of mass'' of the wavepacket).
The phases in Eq.\ref{eq:trans} read
\begin{align}
    \Phi_n = \sum_{j=1}^n \xi_j [\epsilon_0 (t_{2j}-t_{2j-1})+ g(t_{2j})-g(t_{2j-1})],
\label{eq:phin}
\end{align}
where $g(t) = \int_0^t \epsilon_d(t') dt'$ describes the influence of the external drive.
As will become clear below, this is the only contribution of the drive.
The amplitudes, which depend only on the bath, read
\begin{align}
\label{eq:Q}
    Q_n=\exp\left( -\sum_{j=1}^{n} S_{2j, 2j-1} - \sum_{j=2}^n \sum_{k=1}^{j-1} \xi_j \xi_k \Lambda_{j,k}\right) ,
\end{align}
and
\begin{align}
\label{eq:H}
    H_n=\exp\left( i \sum_{j=1}^n \sum_{k=0}^{j-1} \xi_j \eta_k X_{j,k} \right).
\end{align}
The coefficients $\Lambda_{j,k}$ and $X_{j,k}$ appearing above are obtained with the help of 
\begin{align}
    S_{j,k} = S(t_j - t_k),
\end{align}
and
\begin{align}
    R_{j,k} = R(t_j - t_k),
\end{align}
which come from the real and the imaginary parts of the Fourier transform of the bath autocorrelation function,
\begin{align}
    S(\tau) = \frac{q_0^2}{\pi \hbar} \int_0^\infty d\omega \frac{J(\omega)}{\omega} [1-\cos(\omega\tau)] \coth\left(\frac{\beta \hbar\omega}{2}\right),
\end{align}
and
\begin{align}
    R(\tau) = \frac{q_0^2}{\pi \hbar} \int_0^\infty d\omega \frac{J(\omega)}{\omega} \sin(\omega\tau).
\end{align}
Finally, one gets that
\begin{align}
\Lambda_{j,k}=S_{2j,2k-1}+S_{2j-1,2k}-S_{2j,2k}-S_{2j-1,2k-1},
\end{align}
and
\begin{align}
    X_{j,k} = 
    R_{2j,2k+1}+R_{2j-1,2k}-R_{2j,2k}-R_{2j-1,2k+1}.
\end{align}
Note that only $S(\tau)$ (and hence $\Lambda_{j,k}$) depends on the temperature $T = 1/k_B\beta$.

From Eq.\ref{eq:trans} alone, it is not obvious how to identify the work cost of the process.
In what follows, we discuss an appropriate definition of work that allows us to obtain the dissipative adaptation relation from the driven spin-boson model.

\section{Quantum work for the Caldeira-Leggett model within the path-integral formalism}\label{sec:themodynamics}
In equilibrium systems, thermodynamic quantities such as work and heat admit unambiguous definitions, whereas in driven nonequilibrium systems, particularly in the quantum regime, things are more subtle~\cite{zwanzig2001nonequilibrium}.
Recent progress has extended classical stochastic thermodynamics to the quantum domain by characterizing work as a functional of system trajectories within the path integral formalism~\cite{qiu2020path,seifert2012stochastic}. Although quantum trajectories are not directly observable, this approach provides a consistent framework that reduces to classical fluctuating work in the semiclassical limit ($\hbar \rightarrow 0$).

A path-integral formalism for quantum work was developed in Ref. \cite{funo2018path} for systems driven by an arbitrary time-dependent potential $V(\lambda_t, x)$,
\begin{align}
    H_S(t) = \frac{p^2}{2m}+V(\lambda_t,x).
\end{align}
By assuming the so-called two-point measurement (TPM) scheme for the Caldeira-Leggett model, the authors find a quantum work functional, namely 
\begin{align}\label{eq:work}
    W_{\nu}[x] = \int_{0}^{\tau}dt \frac{1}{\hbar \nu} \int_{0}^{\hbar \nu}ds \ \dot{\lambda}_t\dfrac{\partial V[\lambda_t,x(t+s)]}{\partial \lambda_t}.
\end{align}
Although the TPM scheme cannot be directly employed in an experimental setup, it turns out that assuming so gives rise to a theoretically meaningful quantity.
By expanding it in powers of $\nu$ (or $\hbar$), they obtain 
\begin{align}
    W_{\nu}[x]=W_{cl}[x]+\frac{i\nu}{2}W_q^{(1)}[x]+\mathcal{O}(\hbar^2\nu^2).
\end{align}
$W_{cl}[x]$ is the classical work for a particle undergoing quantum fluctuations in its path, $x(t)$.
This reads 
\begin{align}
    W_{cl}[x] = \int_{0}^{\tau}dt \dot{\lambda}_t\dfrac{\partial V[\lambda_t,x(t)]}{\partial \lambda_t}.
    \label{eq:workclass}
\end{align}
The first-order quantum correction is
\begin{align}
    W_q^{(1)}[x] = -i\hbar \int_{0}^{\tau}dt \ \dot{x}(t)\dot{\lambda}_t\frac{\partial^2 V[\lambda_t,x(t)]}{\partial \lambda_t \partial x(t)}.
\end{align}
This path-integral version provides us with a quantum extension of the classical work for a quantum particle classically driven and subjected to a quantized thermal bath at arbitrary temperatures and coupling strengths.

\section{Dissipative adaptation in the driven spin-boson model}
To connect the work functional and the transition probability for the driven spin-boson model, we start from the simplest case.
For a single tunneling event ($n=1$), the transition probability from $\ket{L}$ to $\ket{R}$ reads
\begin{align}
    p^{(n=1)}_{L\rightarrow R}(t) =\frac{\Delta^2}{4}\int_{0}^{t}dt_2\int_{0}^{t_2}dt_1\sum_{\xi_1 = \pm 1 }Q_1 H_1 e^{i\Phi_1},
\label{eq:plrn1}
\end{align}
with the accumulated phase
\begin{align}
    \Phi_1&=\xi_1[\epsilon_0(t_2-t_1)+g(t_2)-g(t_1)]\\
    &=\xi_1\biggr[\epsilon_0(t_2-t_1)+\int_{t_1}^{t_2}dt \epsilon_d(t)\biggl].
\end{align}
We recall that the time-dependent bias enters the system's Hamiltonian as 
$H_S(t)=H_0-\hbar \epsilon_d(t)\sigma_z/2$,
where 
$H_0 = -\hbar\Delta \sigma_x/2 - \hbar \epsilon_0 \sigma_z/2$.
This is the microscopic origin of the external work. 
We expect the rate $\partial_{\epsilon_d} H_S \dot{\epsilon}_d(t) = -\hbar \dot{\epsilon}_d \sigma_z/2$ to be somehow related to the power.
The stochastic nature of the work comes from the uncertainty on the system observable $\sigma_z$.
We recall that $\sigma_z$ here comes from the discretization of the position $q/q_0 = 2(\eta + \xi)$.

In the discretized position formalism, the state of the system is characterized by the random variables $\eta_j$ and $\xi_j$.
However, only $\xi_j$ appears in $\Phi_n$.
Based on the above reasoning, we intuitively propose a work-like function from time $0$ to $t$, along this $n=1$ quantum trajectory, as 
\begin{align}
    W_1(t)\equiv -\int_{0}^{t} \hbar \xi_1 \dot{\epsilon}_d(t')\ dt'.
\end{align}
Without loss of generality, we can always assume that $\epsilon_d(0)=0$, for any deviation is incorporated in the static term $\epsilon_0$.
So the phase can be rewritten as 
\begin{align}
    \Phi_1=\xi_1\epsilon_0(t_2-t_1)-\int_{t_1}^{t_2}  dt \ W_1(t)/\hbar.
\label{eq:phi1w1}
\end{align}
We now have to study the relation between our $W_1$ and the Funo-Quan work $W_{\nu}[x]$ for the driven spin-boson model. 

For a driven two-level system (TLS), the semiclassical work functional $W_{cl}[x]$ reduces to 
\begin{align}
    W^{TLS}_{cl}[x]&= \int_{0}^{t}\dot{\epsilon}_d(t)\;\partial_{\epsilon_d} H_S[\epsilon_d,x(t')]\;dt'\\
    &=-\frac{\hbar}{2} \int_{0}^{t} \sigma_z(t') \dot{\epsilon}_d(t') dt'\\
    &=-\hbar \int_{0}^{t} [\eta(t')+\xi(t')] \dot{\epsilon}_d(t') dt'\\
    &=W^{TLS}_{qs} + W^{TLS}_{ns},
\end{align}
where
\begin{align}
    W_{ns}^{TLS}\equiv -\hbar \int_{0}^{t}\xi(t')\;\dot{\epsilon}_d(t')\;dt'=W_1(t)
\label{eq:wnsw1}
\end{align}
corresponds to the nonstationary contribution to the work due to the buildup of quantum coherence during a short enough time interval $t\rightarrow 0$ so that $\xi(t') \approx \xi_i$.
The other term, namely
$W^{TLS}_{qs} \equiv -\hbar \int_{0}^{t} \eta(t') \dot{\epsilon}_d(t') dt'$, 
quantifies a quasistatic contribution associated with the energy shifts of localized states.
We notice that only the nonstationary work $W^{TLS}_{ns}$ appears in $p^{(n=1)}_{L\rightarrow R}(t)$ (in Eq.(\ref{eq:plrn1})).
In other words, the quasistatic work $W^{TLS}_{qs}$ does not drive coherent transitions.
This is expected, as long as quantum tunneling is the only transition mechanism available in the model.

Equations (\ref{eq:plrn1}), (\ref{eq:phi1w1}), and (\ref{eq:wnsw1}) together establish a connection between the absorbed work and the nonequilibrium population transfer in the light of the dissipative adaptation relation:
the nonstationary work emerges as the nonequilibrium quantity selecting transitions.

We now consider the full expression for 
$p_{L\rightarrow R}(t)$ valid at finite times $t$, 
where arbitrary $n$ jumps must be considered 
(see Eq.(\ref{eq:trans})).
From Eq.(\ref{eq:phin}), we find that
$$g(t_{2j})-g(t_{2j-1}) = -\int_{2j-1}^{2j} dt' W_{ns,j}^{TLS}(t')/\hbar,$$
where
\begin{align}
W_{ns,j}^{TLS}(t')  \equiv -\hbar \xi_j\;\epsilon_d(t')
\end{align}
is the $j$-th component of the nonstationary part of the classical work functional on the two-level system, carried out from time $0$ to time $t'$.
We arrive at our main result, namely
\begin{align}\label{eq:qdasbm}
    p_{L \rightarrow R}(t) &= \sum_{n=1}^\infty (-1)^{n+1}\left(\frac{\Delta}{2}\right)^{2n}\int_{0}^{t}\mathcal{D}_n 
    \left\langle e^{-\frac{i}{\hbar}\int_{t_{2j-1}}^{t_{2j}} {dt'} W_{ns,j}^{TLS}(t')} \right\rangle_n,
\end{align}
where we have defined the bracket symbols as a real-valued generalized average over the trajectories ensemble,
\begin{align}
    \left\langle {\bullet} \right\rangle_n
    \equiv 
    \sum_{\left\{ \xi_j \right\}}
    \sum_{\left\{ \eta_k \right\}'}
    e^{i\xi_j\epsilon_0(t_{2j}-t_{2j-1})}
    Q_n H_n \bullet.
\end{align}
Terms $Q_n$ and $H_n$ have been defined in Eqs.(\ref{eq:Q}) and (\ref{eq:H}).

Under certain circumstances, it is possible to simplify the above expression.
If one assumes that $\Lambda_{j,k} = 0$, for instance, the exponential of the sum factorizes into a product of exponentials, so that
\begin{align}\label{eq:qdasbm2}
    p_{L \rightarrow R}(t) &= \sum_{n=1}^\infty (-1)^{n+1}\left(\frac{\Delta}{2}\right)^{2n}\int_{0}^{t}\mathcal{D}_n 
    \prod_{j=1}^{n}
    \left\{ e^{-\frac{i}{\hbar}\int_{t_{2j-1}}^{t_{2j}} {dt'} W_{ns,j}^{TLS}(t')} \right\}_j,
\end{align}
where now the generalized average reads
$\left\{ {\bullet} \right\}_j \equiv \sum_{\xi_j=-1}^{+1}e^{i\xi_j\epsilon_0(t_{2j}-t_{2j-1})}Q_n^{(j)}H_n^{(j)} \bullet$.
Here,
$Q_n^{(j)} \equiv \exp\left( - S_{2j,2j-1} \right)$,
and
$H_n^{(j)} \equiv \sum_{\left\{ \eta_k = \pm 1 \right\}' } \exp\left(i \xi_j \sum_{k=0}^{j-1} \eta_k X_{j,k} \right)$.
Assuming $\Lambda_{j,k}=0$ is one of the prescriptions within the so-called NIBA (non-interacting blip approximation), valid when the average blip length is much smaller than the average sojourn length \cite{grifoni1995cooperative}.
The NIBA typically becomes exact for strong dampings, or else at high temperatures for weak dampings.
Apart from making $\Lambda_{j,k} = 0$, the other prescription within the NIBA is to set $X_{j,k} = 0$ for all $k\neq j-1$.
In that case, only $\eta_{j-1}X_{j,j-1}$ would survive in the summation inside the exponent of $H_n^{(j)}$.

Equation (\ref{eq:qdasbm}) shows that transition probabilities are explicitly dependent on the exponential of the functional for the absorbed work.
Because this model takes coherent quantum tunneling into account, it generalizes the classical dissipative adaptation relation from Eq.(\ref{eq:england}).
And since the derivation of Eq.(\ref{eq:qdasbm}) is valid for a generic thermal bath (either Markovian or non-Markovian, weakly or strongly coupled, at vanishing or finite temperatures), the present model goes beyond the zero-temperature, Markovian, weakly coupled quantum bath employed in the derivation of Eqs.(\ref{eq:daniel}).

The dissipative adaptation relation for the spin-boson model, Eq.(\ref{eq:qdasbm}), reveals three peculiar features.
First, only the nonstationary term of the work gives a finite contribution.
This makes sense to us, since energy shifts (described by the quasistatic work) cannot induce quantum coherent tunneling.
Second, there appears to be the time integral of the work, not the work itself. 
This is perhaps because the model enables one to describe non-Markovian thermal baths.
Third, the exponential is imaginary. 
This can be regarded as a non-classical signature of the dynamics.
Finally, we note that only the classical part of the work functional is relevant to the present model.
Mathematically, this can be seen from Eq.(\ref{eq:phin}) where $g(t_{2j})-g(t_{2j-1})$ multiplies only $\xi_j$, but not $\xi_{j\pm 1,2,...}$.
The time lag ``$s$'' between the system's position, $x(t+s)$, and the driving force, $\partial_t{\lambda}_t$, is what distinguishes Eq.(\ref{eq:work}) from (\ref{eq:workclass}).

\section{Conclusions} \label{sec:conclusions}
We have revisited the dynamics of the driven spin-boson model, with the aim of elucidating the relationship between nonequilibrium transitions and the work performed by the external drive.
Motivated by the dissipative adaptation hypothesis, proposed by England in 2015, we analyzed whether the probability for a quantum particle trapped in a double well to jump between the two minima could be associated with the work absorbed from the drive and dissipated to the environment.
We have employed the well-known analytical solution to the time-dependent driven spin-boson model. 
We have also assumed a path-integral formulation, based on the two-projective measurements scheme, to compute the work.
We have derived Eq.(\ref{eq:qdasbm}) as our main result. 
This means a generalized version of the dissipative adaptation valid for the driven spin-boson model, where quantum coherent tunneling is taken into account.
The presence of tunneling implies that only the nonstationary work contributes to the transition probability from state $\ket{L}$ to $\ket{R}$.
Our result spans arbitrary system-bath strengths, bath temperatures, and spectral functions.

These findings place the driven spin-boson model at the center of an ongoing debate regarding the relative roles of kinetic effects and thermodynamic principles in nonequilibrium organization.
\cite{mandal2023kinetic, england2015dissipative}. 
Whereas kinetic asymmetry has been shown to generate population inversion and steady-state organization, dissipative adaptation predicts that such configurations should also correlate with enhanced work dissipation \cite{england2015dissipative, valente2021quantum}. 
Our results suggest that further investigation is required to determine which work-related quantities, such as average work, work fluctuations, or exponential work averages, are most relevant when population localization and inversion are present \cite{magazzu2015dissipative,magazzu2018probing,dakhnovskii1995manipulating,goychuk1996control}.
In particular, localization of the population on the energetically unfavorable side of the potential may arise under strong driving without a clear correspondence to increased energy absorption. 
This observation highlights the subtle distinction between nonequilibrium organization driven by kinetic asymmetries and adaptation driven by thermodynamic dissipation \cite{mandal2023kinetic}. 
Establishing a numerical connection between work absorption and stationary localization remains an open problem.

Future studies will focus on clarifying the thermodynamic cost of maintaining nonequilibrium stationary states in driven quantum systems, including the role of quantum coherence, correlations, and entanglement suppression \cite{breuer2002theory}. 
A deeper understanding of dissipative self-organization and adaptation in quantum models may provide valuable information on the emergence of complex behavior in driven systems and contribute to a unified framework connecting quantum thermodynamics, nonequilibrium dynamics, and adaptive behavior \cite{ragazzon2018energy}.

\section*{Acknowledgements} \label{sec:acknowledgements}
The authors acknowledge  Ivan Medina and Sergio R. Muniz for helpful discussions. EIG and RJSA acknowledge support from Coordena\c{c}\~{a}o de Aperfei\c{c}oamento de Pessoal de N\'{i}vel Superior - Brasil (CAPES) - Finance Code 001. DOSP acknowledges support by Brazilian funding agencies CNPq (Grant 307028/2019-4) and FAPESP (Grant No 2017/03727-0). 
DV was supported by the Serrapilheira Institute (Grant No. Serra-1912-32056) and by CNPq (Grant 402074/2023-8) and CNPq (Grant 408990/2025-2).
EIG, RJSA, DOSP and DV acknowledge support from the Instituto Nacional de Ci\^{e}ncia e Tecnologia de Informa\c{c}\~{a}o Qu\^{a}ntica, CNPq INCT- IQ (465469/2014-0), Brazil.

\begin{appendices}

\section{Path integral formalism}\label{appendix:path}
In this appendix, we briefly summarize the path integral formulation used throughout the paper and fix notation. Detailed treatments can be found in Refs.\cite{grifoni1998driven, weiss2012quantum}. 

In classical mechanics, the dynamics of the particle is determined by the principle of minimum action. Given the Lagrangian $L(\dot{x},x,t)$, the action 
\begin{align}
    S[x(t)]=\int_{t_a}^{t_b}L(\dot{x},x,t)\;dt
\end{align}
is extremized by the trajectory, which satisfies the Euler-Lagrange equation
\begin{align*}
    \dfrac{d}{dt}\left(\dfrac{\partial L}{\partial \dot{x}}\right)-\dfrac{\partial L}{\partial x}=0.
\end{align*}
In quantum mechanics, transition amplitudes are obtained by summing over all possible paths connecting an initial point $a = (x_a, t_a)$ to a final point $b=(x_b,t_b)$. The propagator is given by
\begin{align*}
    K(b,a) = \sum_{\mbox{all paths}} \phi[x(t)],
\end{align*}
where each path contributes with a phase
\begin{align}
    \phi[x(t)]=C\;e^{iS[x(t)]/\hbar}
\end{align}
and $S[x(t)]$ is the classical action evaluated along the path.

The sum over paths is defined by discretizing the time interval $[t_a,t_b]$ into $N$ steps of width $\epsilon = (t_b-t_a)/N$, introducing intermediate coordinates $\{x_i\}$, and integrating over all possible configurations. In the continuum limit $\epsilon \rightarrow 0$, the propagator takes the form 
\begin{align}
    K(b,a)=\int_{x(t_a) = x_a}^{x(t_b)=x_b} \mathcal{D}x(t) e^{(i/\hbar)S[x(t)]},
\end{align}
where $\mathcal{D}x(t)$ denotes the path integral measure, defined up to a normalization factor that depends on the specific Lagrangian. 

This formulation provides a natural framework for treating open quantum systems, where environmental degrees of freedom can be integrated out exactly, leading to an effective description in terms of influence functionals. In the main text, this formalism is employed to derive transition probabilities and to establish the connection between driven dynamics, dissipation, and quantum work. 
\section{Functional of work}\label{appendix:workfunctional}
The derivation follows Refs.\cite{funo2018path,funo2018patheat,qiu2020path}. We briefly outline the steps leading to the path integral representation of the characteristic function of work. 

The characteristic function is defined as 
\begin{align}\label{eq:1}
    \chi_{W}(\nu)=\mbox{Tr}[U_Se^{-i\nu H_S(\lambda_0)}\rho_S(0)U^{\dagger}_{S}e^{i\nu H_S(\lambda_{\tau})}],
\end{align}
where $U(\tau) = \mathcal{T}[\exp(-i\hbar \int_{0}^{\tau}dtH(\lambda_{\tau}))]$ is the time evolution operator, $[0,\tau]$ is the time interval for controlled parameter $\lambda_{\tau}$, and
\begin{align}
    H_S=p^2/2m+V(\lambda_{\tau},x),
\end{align}
is the Hamiltonian, and $\rho_S(0)$ is the initial state of the system.

Using coordinate resolutions of the identity and standard path integral representations for the forward and backward propagators
\begin{align}
    \expval{x_f|U_Se^{-i\nu H_S(\lambda_0)}|x_i} &= \int \mathcal{D}x e^{(i/\hbar)S_{1}^{\nu}[x]},\\
    \expval{y_i|U_S^{\dagger}e^{i\nu H_S(\lambda_{\tau})}|y_f} &= \int \mathcal{D}y e^{(-i/\hbar)S_{2}^{\nu}[y]}, \\
    \int dx \ket{x}\bra{x}&=1
\end{align}
Eq.(\ref{eq:1}) can be written as 
\begin{align}
    \chi_{W}(\nu)=&\int dx_idx_i^{'}dx_fdx_f^{'}\delta(x_f-x_f^{'})\mathcal{D}x\mathcal{D}x'\\
    &e^{(i/\hbar)(S_{1}^{\tau}[x]-S_{2}^{\tau}[x'])}\rho(x_i,x_i^{'}),
\end{align}
where the modified actions are
\begin{align}
    S_{1}^{\tau}[x]&= \int_{0}^{\hbar\nu}dt\mathcal{L}[\lambda_0,x(t)]+\int_{\hbar \nu}^{\tau+\hbar\nu}dt\mathcal{L}[\lambda_{t-\hbar \nu},x(t)],\nonumber \\
    S_{2}^{\tau}[x]&= \int_{0}^{\tau}ds\mathcal{L}[\lambda_{s},x'(s)]+\int_{\tau}^{\tau+\hbar\nu}ds\mathcal{L}[\lambda_{\tau},x'(s)].
\end{align}

A key identity allows the separation of a work-dependent contribution, 
\begin{align}
    \frac{i}{\hbar}S_{1}^{\tau}[x]=\frac{i}{\hbar}S_{2}^{\tau}[x]+i\nu W_{\nu}[x].
\end{align}
Taking derivatives with respect to the upper time boundary and integrating over the protocol duration yields
\begin{equation}
S_1^\tau[x]-S_2^\tau[x]
=
-\int_0^\tau dt \int_t^{t+\hbar\nu} ds,
\dot{\lambda}_t
\frac{\delta \mathcal{L}[\lambda_t,x(s)]}{\delta \lambda_t}.
\end{equation}
The work functional is therefore identified as 
\begin{align}
    W_{\nu}[x,\tau] =\frac{1}{\hbar \nu}\int_{0}^{\tau}dt\int_{0}^{\hbar \nu}ds \dot{\lambda}_{t}\frac{\delta \mathcal{L}[\lambda_t,x(s+t)]}{\delta \lambda_t}.
\end{align}
In the limit $\nu \rightarrow 0$, this expression reduces to the standard classical definition of work performed along a trajectory.
\end{appendices}
\bibliography{sn-bibliography}
\end{document}